\documentclass[12pt]{article}
\usepackage{amssymb}
\usepackage{graphicx}
\usepackage{lscape}
\renewcommand{\theequation}{\arabic{section}.\arabic{equation}}

\begin{document}



\def\a{\alpha}
\def\b{\beta}
\def\d{\delta}
\def\e{\epsilon}
\def\g{\gamma}
\def\h{\mathfrak{h}}
\def\k{\kappa}
\def\l{\lambda}
\def\o{\omega}
\def\p{\wp}
\def\r{\rho}
\def\t{\tau}
\def\s{\sigma}
\def\z{\zeta}
\def\x{\xi}
\def\V={{{\bf\rm{V}}}}
 \def\A{{\cal{A}}}
 \def\B{{\cal{B}}}
 \def\C{{\cal{C}}}
 \def\D{{\cal{D}}}
\def\G{\Gamma}
\def\K{{\cal{K}}}
\def\O{\Omega}
\def\R{\bar{R}}
\def\T{{\cal{T}}}
\def\L{\Lambda}
\def\f{E_{\tau,\eta}(sl_2)}
\def\E{E_{\tau,\eta}(sl_n)}
\def\Zb{\mathbb{Z}}
\def\Cb{\mathbb{C}}

\def\R{\overline{R}}

\def\beq{\begin{equation}}
\def\eeq{\end{equation}}
\def\bea{\begin{eqnarray}}
\def\eea{\end{eqnarray}}
\def\ba{\begin{array}}
\def\ea{\end{array}}
\def\no{\nonumber}
\def\le{\langle}
\def\re{\rangle}
\def\lt{\left}
\def\rt{\right}

\newtheorem{Theorem}{Theorem}
\newtheorem{Definition}{Definition}
\newtheorem{Proposition}{Proposition}
\newtheorem{Lemma}{Lemma}
\newtheorem{Corollary}{Corollary}
\newcommand{\proof}[1]{{\bf Proof. }
        #1\begin{flushright}$\Box$\end{flushright}}

\baselineskip=20pt

\newfont{\elevenmib}{cmmib10 scaled\magstep1}
\newcommand{\preprint}{
   \begin{flushleft}
   \end{flushleft}\vspace{-1.3cm}
   \begin{flushright}\normalsize
   \end{flushright}}
\newcommand{\Title}[1]{{\baselineskip=26pt
   \begin{center} \Large \bf #1 \\ \ \\ \end{center}}}
\newcommand{\Author}{\begin{center}
   \large \bf

Xiaotian Xu${}^a$, Junpeng Cao${}^{b,c}$,~Shuai Cui${}^{b}$,~Wen-Li Yang${}^{a,d}\footnote{Corresponding author:
wlyang@nwu.edu.cn}$,~Kangjie Shi${}^a$ and~Yupeng
Wang${}^{b,c}$
 \end{center}}
\newcommand{\Address}{\begin{center}

     ${}^a$Institute of Modern Physics, Northwest University,
     Xian 710069, China\\
     ${}^b$Beijing National Laboratory for Condensed Matter
           Physics, Institute of Physics, Chinese Academy of Sciences, Beijing
           100190, China\\
     ${}^c$Collaborative Innovation Center of Quantum Matter, Beijing,
     China\\
     ${}^d$Beijing Center for Mathematics and Information Interdisciplinary Sciences, Beijing, 100048,  China
   \end{center}}
\newcommand{\Accepted}[1]{\begin{center}
   {\large \sf #1}\\ \vspace{1mm}{\small \sf Accepted for Publication}
   \end{center}}

\preprint
\thispagestyle{empty}
\bigskip\bigskip\bigskip

\Title{Off-diagonal Bethe Ansatz solution of the $\tau_2$-model} \Author

\Address
\vspace{1cm}

\begin{abstract}
The generic quantum $\tau_2$-model
(also known as Baxter-Bazhanov-Stroganov (BBS) model) with periodic boundary condition is studied via the off-diagonal Bethe Ansatz method.
The eigenvalues of the corresponding
transfer matrix (solutions of the recursive functional relations in $\tau_j$-hierarchy) with generic site-dependent inhomogeneity parameters  are given in terms of an inhomogeneous $T-Q$ relation with polynomial $Q$-functions. The associated
Bethe Ansatz equations are obtained. Numerical solutions of the Bethe Ansatz equations for small number of sites indicate that the inhomogeneous $T-Q$ relation does indeed give the complete spectrum.

\vspace{1truecm} \noindent {\it PACS:} 75.10.Pq, 02.30.Ik, 71.10.Pm

\end{abstract}
\newpage
\section{Introduction}
\label{intro} \setcounter{equation}{0}
Among quantum integrable
models, the $\tau_2$ (BBS)-model \cite{Baz90} plays a special role
for its unique properties, e.g., it is one of the simplest quantum
integrable models associated with cyclic representation of the Weyl
algebra; it allows to include multiple inhomogeneity parameters on
each single site without breaking the integrability of the model;
and more interestingly, the $\tau_2$-model under certain parameter
constraint is highly related to some other integrable models such as
the chiral Potts model \cite{vG,alc,p,perk1,perk2,37} and the
relativistic quantum Toda chain model \cite{toda}. Many papers have
appeared in literature for such connections and many efforts have
been made to obtain the solutions of chiral Potts model by solving
the $\tau_2$-model with a recursive functional relation
\cite{45,43,49,39}. However, it was found that only in the
super-integrable sub-sector \cite{vG} the algebraic Bethe Ansatz
method can be applied on this model to obtain Baxter's $T-Q$
\cite{Bax82} solutions and  Bethe Ansatz equations, while for the
generic $\tau_2$-model, though its integrability \cite{Baz90} was
proven, there is no simple $Q$-operator solution in terms of
Baxter's $T-Q$ relation. The $Q$-operator is in fact a very
complicated function defined in high genus space and its concrete
form is still hard to be derived.

In this paper, we adopt the off-diagonal Bethe Ansatz method \cite{cao1} (for comprehensive introduction, see \cite{Wan15}) to study the quantum $\tau_2$-model. It seems that the situation of the generic $\tau_2$-model is quite similar to the quantum XYZ model with an odd number of sites \cite{cao1,Wan15}, in which there is also no simple polynomial solutions of the $Q$-function in terms of Baxter's $T-Q$ relation. However, by including an extra off-diagonal term in the $T-Q$ relation (i.e., the inhomogeneous $T-Q$ relation), we show that the eigenvalues of the generic $\tau_2$ transfer matrix can be expressed explicitly in terms of a trigonometric polynomial $Q$ function and thus a proper set of Bethe Ansatz equations can be derived.

The structure of the paper is the following. In the subsequent section, we give a brief introduction of the $\tau_2$ transfer matrix. In section 3, we study the fundamental properties of the transfer matrix and its fusion hierarchy. In section 4, we give the eigenvalues of the transfer matrix and the associated Bethe Ansatz equations. Concluding remarks are given in section 5 and the detailed proofs about the inhomogeneous $T-Q$ relation and its degenerate case are given in Appendices $A$ \& $B$.


\section{ Transfer matrix}
\label{XXZ} \setcounter{equation}{0}

Let $R(u)\in {\rm End}({\rm\bf C^2}\otimes {\rm\bf C^2})$ be the six-vertex $R$-matrix
\bea
R(u)=\left(
\begin{array}{cccc}
  \sinh(u+\eta) & 0 & 0 & 0 \\
  0 & \sinh u & \sinh \eta & 0 \\
  0 & \sinh \eta & \sinh u & 0 \\
  0 & 0 & 0 & \sinh(u+\eta)
\end{array}\right),\label{R-matrix}
\eea
with the crossing parameter $\eta$ taking the special values \footnote{It corresponds to the case that
$q=e^{-\eta}$ is a $p$-root of unity: $q^p=1$. The generalization to the case of $\eta=2i\pi p'/p$ with two
coprime positive integers $p'$ and $p$ is straightforward.}:
\bea
\eta=2i\pi /p,\quad p=2l+1,\quad l=1,2,\cdots.\label{root}
\eea
The $R$-matrix satisfies
the quantum Yang-Baxter equation (QYBE) \cite{Bax82,Kor93} and has played an important role in the
quantum integrable systems and the quantum group theories \cite{Cha94}. Moreover, the $R$-matrix becomes
some projectors when the spectral parameter $u$ takes some special values:
\bea
&&\mbox{Antisymmetric-fusion conditions}:\,R(-\eta)=-2\sinh\eta P^{(-)},\label{Fusion-Con-1}\\[6pt]
&& \mbox{Symmetric-fusion conditions}:\, R(\eta) = 2\sinh\eta\,{\rm
Diag}(\cosh\eta,1,1,\cosh\eta)\,P^{(+)},\label{Fusion-Con-2} \eea
where $P^{(+)}$ ($P^{(-)}$) is the symmetric (anti-symmetric)
projector of the tensor space ${\rm\bf C^2}\otimes {\rm\bf C^2}$.

Let ${\rm\bf V}$ denote a $p$-dimensional linear space (i.e. the local Hilbert space)
with an orthonormal basis $\{|m\rangle~|m\in \Zb_p\}$.
$X$ and $Z$ are two $p\times p$ matrices acting on the basis as follows:
\bea
X|m\rangle =q^m|m\rangle,\quad Z|m\rangle=|m+1\rangle,\quad q=e^{-\eta},\quad m\in \Zb_p. \label{XZ-actions}
\eea
Here and below we adopt the standard
notations: for any matrix $A\in {\rm End}({\rm\bf V})$, $A_n$ is an
embedding operator in the tensor space ${\rm\bf V}\otimes
{\rm\bf V}\otimes\cdots$, which acts as $A$ on the $n$-th space and as
identity on the other factor spaces. Then the embedding operators $\{X_n,Z_n|n=1,\cdots,N\}$ satisfy the ultra-local Weyl algebra:
\bea
X_nZ_m=q^{\delta_{nm}}Z_mX_n,\quad X_n^p=Z_n^p=1,~~\forall n,m\in\{1,\cdots,N\}.\label{Weyl}
\eea

The $\tau_2$-model can be described by an quantum spin chain \cite{Baz90}. With each site $n$ of the
quantum chain, the associated $L$-operator $L_n(u)\in {\rm End}({\rm \bf C^2}\otimes{\rm\bf V})$ defined in
the most general cyclic representation of $U_q(sl_2)$,  is given by \cite{Baz90}
\bea
L_{n}(u)&=&\left(
\begin{array}{cc}
  e^{u}d^{(+)}_nX_n+e^{-u}d^{(-)}_nX^{-1}_n & (g^{(+)}_nX^{-1}_n+g^{(-)}_nX_n)Z_n \\[6pt]
  (h^{(+)}_nX^{-1}_n+h^{(-)}_nX_n)Z^{-1}_n & e^{u}f^{(+)}_nX^{-1}_n+e^{-u}f^{(-)}_nX_n
\end{array}\right)\no\\[6pt]
&=&\left(\begin{array}{cc}
                           A_n(u) & B_n(u) \\[6pt]
                           C_n(u) & D_n(u)
                         \end{array}
\right),\quad n=1,\ldots, N,\label{L-operator} \eea where
$d^{(+)}_n$, $d^{(-)}_n$, $g^{(+)}_n$, $g^{(-)}_n$, $h^{(+)}_n$,
$h^{(-)}_n$, $f^{(+)}_n$ and $f^{(-)}_n$ are some parameters
associated with  the  $n$-th site. These parameters are subjected to two
constraints: \bea g^{(-)}_nh^{(-)}_n=f^{(-)}_nd^{(+)}_n,\quad
g^{(+)}_nh^{(+)}_n=f^{(+)}_nd^{(-)}_n,\quad
n=1,\cdots,N.\label{Constraint-1} \eea It was shown \cite{Baz90}
that the $L$-operators satisfy the relations: \bea
R(u-v)(L_{n}(u)\otimes 1)(1\otimes L_{n}(v))=(1\otimes
L_{n}(v))(L_{n}(u)\otimes 1)R(u-v),\quad n=1,\ldots,N,
\label{Rll-relation} \eea where the $R$-matrix $R(u)$ is given by
(\ref{R-matrix}). The corresponding one-row monodromy  matrix $T(u)$
is thus defined as: \bea T(u)= \left(\begin{array}{cc}
                           {\bf A}(u) & {\bf B}(u) \\[6pt]
                           {\bf C}(u) & {\bf D}(u)
                         \end{array}
\right)=L_N(u)\,L_{N-1}(u)\,\cdots L_1(u),\label{Mono-T}
\eea which satisfies the quadratic relation known as the Yang-Baxter algebra
\bea
R(u-v)(T(u)\otimes 1)(1\otimes T(v))=(1\otimes T(v))(T(u)\otimes 1)R(u-v).\label{RTT-relation}
\eea The transfer matrix $t(u)$ of the $\tau_2$-model with periodic boundary condition is then
given by the partial trace of the monodromy matrix $T(u)$ in the auxiliary space, namely,
\bea
t(u)=tr\lt(T(u)\rt)= {\bf A}(u)+{\bf D}(u).\label{transfer-matrix}
\eea
The quadratic relation (\ref{RTT-relation}) leads to the fact that the transfer matrices with different spectral parameters
are mutually commutative \cite{Kor93}, i.e., $[t(u), t(v)] =0$, which guarantees the integrability of the model
by treating $t(u)$ as the generating functional of the conserved quantities.

The aim of this paper is to construct the eigenvalues $\L(u)$ of the transfer matrix $t(u)$ for generic inhomogeneity parameters
$\{d^{(\pm)}_n,~f^{(\pm)}_n,~g^{(\pm)}_n,~h^{(\pm)}_n |n=1,\cdots,N\}$ obeying the constraints (\ref{Constraint-1}).


\section{ Properties of the transfer matrix}
\label{Transfer} \setcounter{equation}{0}

\subsection{Asymptotic behaviors and average values}
Following \cite{Bax04, Geh06}, let us introduce the operator
${\cal{Q}}$ which commutes with the transfer matrix \bea
{\cal{Q}}=\prod_{n=1}^NX_n, \quad [{\cal{Q}},t(u)]=0,\quad
{\cal{Q}}^p={\rm id}.\label{Q-operator} \eea The explicit expression
(\ref{L-operator}) of the $L$-operator and the definition
(\ref{Mono-T}) of the monodromy matrix $T(u)$ imply that the
transfer matrix $t(u)$ given by (\ref{transfer-matrix}) enjoys the
asymptotic behavior:
\bea
\lim_{u\rightarrow \pm\infty}t(u)=e^{\pm
Nu}
     \lt\{D^{(\pm)}\,{\cal{Q}}^{\pm 1}+F^{(\pm)}\,{\cal{Q}}^{\mp 1}\rt\}+\cdots,\label{Asymp}
\eea
where $D^{(\pm)}$ and $F^{(\pm)}$ are four constants related to the inhomogeneous parameters as follows:
\bea
D^{(\pm)}=\prod_{n=1}^Nd^{(\pm)}_n,\quad F^{(\pm)}=\prod_{n=1}^Nf^{(\pm)}_n.\label{Constant-1}
\eea

Moreover, (\ref{L-operator}) allows us to derive the quasi-periodicity
\bea
&&L_{n}(u+i\pi)=-\s^z\,L_{n}(u)\,\s^z, \label{L-quasi-periodic}
\eea which leads to the quasi-periodicity of the transfer matrix $t(u)$
\bea
t(u+i\pi)=(-1)^Nt(u).\label{transfer-periodic}
\eea The above relation implies that the transfer matrix $t(u)$ can be expressed in terms of $e^u$ as a Laurent polynomial of the form
\bea
t(u)=e^{Nu}t_N+e^{(N-2)u}t_{N-1}+\cdots+e^{-Nu}t_0,\label{Expansion-1}
\eea where $\{t_n|n=0,1\cdots,N\}$ form the $N+1$ conserved charges. In particular, $t_N$ and $t_0$ are
given by
\bea
&&t_N=D^{(+)}\,{\cal{Q}}+F^{(+)}\,{\cal{Q}}^{-1},\no\\
&&t_0=D^{(-)}\,{\cal{Q}}^{-1}+F^{(-)}\,{\cal{Q}},\no
\eea where the  constants $D^{(\pm)}$ and $F^{(\pm)}$ are given by (\ref{Constant-1}).

The property (\ref{Fusion-Con-1}) of the $R$-matrix and the relation (\ref{RTT-relation}) enables one to
introduce the quantum determinant\cite {Kul81,Ize81} of the associated Yang-Baxter algebra
\bea
{\rm Det}_q\lt(T(u)\rt)={\bf A}(u){\bf D}(u-\eta)-{\bf B}(u){\bf C}(u-\eta).
\eea Direct calculation shows that it is proportional to the identity operator and  has the factorized form:
\bea
 &&{\rm Det}_q\lt(T(u)\rt)=\prod_{n=1}^N\,{\rm Det}_q\lt(L_n(u)\rt)=a(u)d(u-\eta)\times{\rm id}
 \stackrel{{\rm def}}{=} \d(u)\times{\rm id},\label{deta-function}\\[6pt]
 &&a(u)=e^{-\frac{N}{2}\eta}\,\lt\{D^{(+)}F^{(+)}\rt\}^{\frac{1}{2}}\,
         \prod_{n=1}^N\lt(e^{u+\eta}-e^{-u-\eta}e^{2\eta}
         \frac{g^{(-)}_nh^{(+)}_n}{d^{(+)}_nf^{(+)}_n}\rt),\label{a-function}\\[6pt]
 &&d(u)=e^{-\frac{N}{2}\eta}\,\lt\{D^{(+)}F^{(+)}\rt\}^{\frac{1}{2}}\,
    \prod_{n=1}^N\lt(e^{u}-e^{-u}
    \frac{g^{(+)}_nh^{(-)}_n}{d^{(+)}_nf^{(+)}_n}\rt),\label{d-function}
\eea where  $D^{(\pm)}$ and $F^{(\pm)}$ are given by (\ref{Constant-1}).

Let us define the average value ${\cal{O}}(u)$ of the matrix elements of the monodromy matrix $T(u)$ (or the
$L$-operators $L_n(u)$)  using the averaging procedure \cite{Tar92}:
\bea
 {\cal{O}}(u)=\prod_{m=1}^p\,O(u-m\eta),
\eea where the operator $O(u)$ can be $\{{\bf A}(u),\,{\bf
B}(u),\,{\bf C}(u),\,{\bf D}(u)\}$ or $\{A_n(u), B_n(u), C_n(u),
D_n(u)$ $| n=1, \cdots, N\}$. It was shown \cite{Tar92} that \bea
&&{\cal{T}}(u)=\left(\begin{array}{cc}
                           {\bf {\cal{A}}}(u) & {\bf {\cal{B}}}(u) \\
                           {\bf {\cal{C}}}(u) & {\bf {\cal{D}}}(u)
                         \end{array}
\right)={\cal{L}}_N(u)\,{\cal{L}}_{N-1}(u)\,\cdots {\cal{L}}_1(u),\label{Average-1}
\eea
where the average value of each $L$-operator is given by
\bea
 {\cal{L}}_n(u)&=&\left(\begin{array}{cc}
                           {{\cal{A}}}_n(u) & { {\cal{B}}}_n(u) \\[6pt]
                           {{\cal{C}}}_n(u) & {{\cal{D}}}_n(u)
                         \end{array}
\right)\no\\[6pt]
&=&
\left(
\begin{array}{cc}
  e^{pu}\{d^{(+)}_n\}^p+e^{-pu}\{d^{(-)}_n\}^p & \{g^{(+)}_n\}^p+\{g^{(-)}_n\}^p \\[6pt]
  \{h^{(+)}_n\}^p+\{h^{(-)}_n\}^p & e^{pu}\{f^{(+)}_n\}^p+e^{-pu}\{f^{(-)}_n\}^p
\end{array}\right), \label{Average-2}
\eea and $n=1, \cdots, N$. It is remarked that the average values of
the matrix elements are Laurent polynomials of $e^{pu}$, which
implies \bea
&&{\cal{T}}(u+\eta)={\cal{T}}(u),\quad {\cal{L}}_n(u+\eta)={\cal{L}}_n(u),\quad n=1,\cdots,N, \label{P-periodic}\\
&&\lim_{u\rightarrow\pm \infty}{\bf {\cal{A}}}(u)=e^{\pm pNu}\lt\{D^{(\pm )}\rt\}^p,\label{A-asymp}\\
&&\lim_{u\rightarrow\pm \infty}{\bf {\cal{D}}}(u)=e^{\pm pNu}\lt\{F^{(\pm )}\rt\}^p,\label{D-asymp}
\eea where the  constants $D^{(\pm)}$ and $F^{(\pm)}$ are given by (\ref{Constant-1}).

\subsection{Fusion hierarchy and truncation identity}

The transfer matrix $t(u)$ given by (\ref{transfer-matrix}) is constructed by tracing over a spin-$\frac{1}{2}$ (i.e.,
two-dimensional) auxiliary space. Using the fusion procedure \cite{Kul81,Kul82,Kir87}, the arbitrary  higher spin-$j$
($j=1,\frac{3}{2},2\cdots$) transfer matrices $t^{(j)}(u)$ which correspond to  spin-$j$ auxiliary spaces and the
same quantum space, i.e., the $N$-tensor space ${\rm\bf V}\otimes
{\rm\bf V}\otimes\cdots$ can be constructed. These transfer matrices $\{t^{(j)}(u)|j=\frac{1}{2},1,\frac{3}{2},2\cdots\}$ (including the transfer
matrix $t(u)$ given by (\ref{transfer-matrix}) as the first one: $t(u)= t^{(\frac{1}{2})}(u)$) commute with each other
\bea
[t^{(j)}(u),\,t^{(j')}(v)]=0,\quad j,j'\in \frac
12, 1, \frac{3}{2}, \cdots,\label{Commut}
\eea
 and obey the fusion
hierarchy relations \cite{Kul82,Kir87,Baz90,Bax04}
\bea
 t^{(\frac{1}{2})}(u)\, t^{(j-\frac{1}{2})}(u- j\eta)&=&
t^{(j)}(u-(j-\frac{1}{2})\eta) + \delta(u)\,
t^{(j-1)}(u-(j+\frac{1}{2})\eta),\no\\[6pt]
j &=&\frac 12, 1, \frac{3}{2}, \cdots, \label{Hier-1} \eea where we
have used the conventions $t^{(-\frac{1}{2})}(u)=0$ and
$t^{(0)}={\rm id}$. The coefficient function $\delta(u)$ related to
the quantum determinant is given by (\ref{deta-function}). Similar
higher-order functional relations have been obtained for RSOS models
\cite{Bax82,Bax82-1, Baz89} and for the 8-vertex model \cite{Bax89}.
Using the recursive relation (\ref{Hier-1}), we can express the
fused transfer matrix $t^{(j)}(u)$ in terms of the fundamental one
$t^{(\frac{1}{2})}(u)$ with a $2j$-order functional relation which
can be expressed as the determinant of some $2j\times 2j$ matrix
\cite{Baz89}, namely, {\small \bea
&&\hspace{-1.4truecm}t^{(j)}(u)\hspace{-0.092truecm}=\hspace{-0.092truecm}\lt|\hspace{-0.12truecm}
\begin{array}{ccccc}
                           t(u\hspace{-0.08truecm}+\hspace{-0.08truecm}(j\hspace{-0.08truecm}-\hspace{-0.08truecm}\frac{1}{2})\eta)
                           &\hspace{-0.08truecm}-\hspace{-0.08truecm}a(u\hspace{-0.08truecm}+\hspace{-0.08truecm}(j\hspace{-0.08truecm}-\hspace{-0.08truecm}\frac{1}{2})\eta)
                           &&&\\[6pt]
                           \hspace{-0.08truecm}-\hspace{-0.08truecm}d(u\hspace{-0.08truecm}+\hspace{-0.08truecm}(j\hspace{-0.08truecm}-\hspace{-0.08truecm}\frac{3}{2})\eta) &t(u\hspace{-0.08truecm}+\hspace{-0.08truecm}(j\hspace{-0.08truecm}-\hspace{-0.08truecm}\frac{3}{2})\eta)
                           &\hspace{-0.08truecm}-\hspace{-0.08truecm}a(u\hspace{-0.08truecm}+\hspace{-0.08truecm}(j\hspace{-0.08truecm}-\hspace{-0.08truecm}\frac{3}{2})\eta)
                           &&\\[6pt]
                           &\ddots&&&\\[6pt]
                           &&\cdots&&\\[6pt]
                           &&&\ddots&\\[6pt]
                           &&\hspace{-0.08truecm}-\hspace{-0.08truecm}d(u\hspace{-0.08truecm}-\hspace{-0.08truecm}(j\hspace{-0.08truecm}+\hspace{-0.08truecm}\frac{1}{2})\eta)
                           &t(u\hspace{-0.08truecm}-\hspace{-0.08truecm}(j\hspace{-0.08truecm}+\hspace{-0.08truecm}\frac{1}{2})\eta)
                           &\hspace{-0.08truecm}-\hspace{-0.08truecm}a(u\hspace{-0.08truecm}-\hspace{-0.08truecm}(j\hspace{-0.08truecm}+\hspace{-0.08truecm}\frac{1}{2})\eta)\\[4pt]
                           &&&\hspace{-0.08truecm}-\hspace{-0.08truecm}d(u\hspace{-0.08truecm}-\hspace{-0.08truecm}(j\hspace{-0.08truecm}-\hspace{-0.08truecm}\frac{1}{2})\eta)
                           &t(u\hspace{-0.08truecm}-\hspace{-0.08truecm}(j\hspace{-0.08truecm}-\hspace{-0.08truecm}\frac{1}{2})\eta)
                         \end{array}
\hspace{-0.12truecm}\rt|,\no\\[8pt]
&&\quad\quad j=\frac{1}{2},1,\frac{3}{2},\cdots,\label{Dert-rep}
\eea} where the functions $a(u)$ and $d(u)$ are given by
(\ref{a-function}) and (\ref{d-function}).

When the crossing parameter $\eta$ takes the special values (\ref{root}), which correspond
to the case of the root of unity, the
spin-$\frac{p}{2}$ transfer matrix satisfy the truncation identity \cite{Baz90,Tar92,Bax04}
\bea
t^{(\frac{p}{2})}(u)=\lt({\bf {\cal{A}}}(u)+ {\bf {\cal{D}}}(u)\rt)\times{\rm id}
       +\d(u-(\frac{p-1}{2})\eta)t^{(\frac{p-2}{2})}(u),\label{Truncation-oper}
\eea  where the functions ${\bf {\cal{A}}}(u)$ and ${\bf {\cal{D}}}(u)$ are the average values of the operators ${\bf A}(u)$ and
${\bf D}(u)$, and are given by (\ref{Average-1})-(\ref{Average-2}). It is remarked that $\frac{p-1}{2}$ is an integer and the functions
${\bf {\cal{A}}}(u)$ and ${\bf {\cal{D}}}(u)$ are invariant  under shifting with $\eta$ (\ref{P-periodic}).

In the following part of the paper, we shall show that the asymptotic behaviors (\ref{Asymp}), the determinant representation (\ref{Dert-rep})
of the transfer matrix $t^{(\frac{p}{2})}(u)$ and the truncation identity (\ref{Truncation-oper}) completely determine the eigenvalues
of the fundamental transfer matrix $t(u)$ given by (\ref{transfer-matrix}). Then with the help of (\ref{Dert-rep}) we can obtain eigenvalues
of all the others higher spin-$j$ transfer matrices $t^{(j)}(u)$.


\section{ Eigenvalues of the fundamental transfer matrix}
\label{Eigenvalues} \setcounter{equation}{0}

\subsection{Functional relations of eigenvalues}
The commutativity (\ref{Commut}) of the fused transfer matrices $\{t^{(j)}(u)\}$ with different spectral
parameters implies that they have common eigenstates. Let $|\Psi\rangle$  be a common eigenstate of these
fused transfer matrices with the eigenvalues $\Lambda^{(j)}(u)$
\bea
t^{(j)}(u)|\Psi\rangle =\Lambda^{(j)}(u)|\Psi\rangle.\no
\eea
The relation (\ref{Q-operator}) allows us to decompose the whole Hilbert space $\cal{H}$ into $p$ subspaces, i.e., ${\cal{H}}= \oplus _{k\in\Zb_p}{\cal{H}}^{(k)}$ according to the action of the operator $ {\cal{Q}}$:
\bea
{\cal{Q}}\,{\cal{H}}^{(k)}=q^k \,{\cal{H}}^{(k)},\quad k\in\Zb_p.
\eea
The commutativity of the transfer matrices and the
operator $ {\cal{Q}}$  implies that each of the subspace is invariant under $t^{(j)}(u)$. Hence the whole set of eigenvalues of the
transfer matrices can be decomposed into $p$ series, denoted by $\L^{(j)}_k(u)$ respectively. The eigenstates corresponding to $\L^{(j)}_{k}(u)$ belong to the subspace ${\cal{H}}^{(k)}$.

The quasi-periodicity (\ref{transfer-periodic}) of the transfer
matrix $t(u)$ implies that the corresponding eigenvalue $\L_k(u)$
satisfies the property \bea
\L_k(u+i\pi)=(-1)^N\L_k(u).\label{Eigen-periodic} \eea The
asymptotic behavior (\ref{Asymp}) of the transfer matrix $t(u)$
gives rise to the fact that the corresponding eigenvalue $\L_k(u)$ enjoys the
behavior \bea \lim_{u\rightarrow \pm\infty}\L_k(u)=e^{\pm Nu}
     \lt\{q^{\pm k}D^{(\pm)}+q^{\mp k}F^{(\pm)}\rt\}+\cdots.\label{Eigen-Asymp}
\eea The analyticity of the $L$-operator (\ref{L-operator}), the
quasi-periodicity (\ref{Eigen-periodic}) and (\ref{Eigen-Asymp})
imply that the eigenvalue $\Lambda_k(u)$ possesses the following
analytical property \bea \L_k(u) \mbox{, as a function of $e^u$, is
a Laurent polynomial of degree $N$ like}~
(\ref{Expansion-1}).\label{Eigenvalue-Anal} \eea

The fusion hierarchy relation (\ref{Hier-1}) and the determinant
representation (\ref{Dert-rep}) of the fused transfer matrices
allows one to express all the eigenvalues $\Lambda^{(j)}_k(u)$ in
terms of the fundamental one $\L_k(u)=\Lambda_k^{(\frac{1}{2})}(u)$
by {\small \bea
&&\hspace{-1.4truecm}\L_k^{(j)}(u)\hspace{-0.092truecm}=\hspace{-0.092truecm}\lt|\hspace{-0.12truecm}
\begin{array}{ccccc}
                           \L_k(u\hspace{-0.08truecm}+\hspace{-0.08truecm}(j\hspace{-0.08truecm}-\hspace{-0.08truecm}\frac{1}{2})\eta)
                           &\hspace{-0.08truecm}-\hspace{-0.08truecm}a(u\hspace{-0.08truecm}+\hspace{-0.08truecm}(j\hspace{-0.08truecm}-\hspace{-0.08truecm}\frac{1}{2})\eta)
                           &&&\\[6pt]
                           \hspace{-0.08truecm}-\hspace{-0.08truecm}d(u\hspace{-0.08truecm}+\hspace{-0.08truecm}(j\hspace{-0.08truecm}-\hspace{-0.08truecm}\frac{3}{2})\eta) &\L_k(u\hspace{-0.08truecm}+\hspace{-0.08truecm}(j\hspace{-0.08truecm}-\hspace{-0.08truecm}\frac{3}{2})\eta)
                           &\hspace{-0.08truecm}-\hspace{-0.08truecm}a(u\hspace{-0.08truecm}+\hspace{-0.08truecm}(j\hspace{-0.08truecm}-\hspace{-0.08truecm}\frac{3}{2})\eta)
                           &&\\[6pt]
                           &\ddots&&&\\[6pt]
                           &&\cdots&&\\[6pt]
                           &&&\ddots&\\[6pt]
                           &&\hspace{-0.08truecm}-\hspace{-0.08truecm}d(u\hspace{-0.08truecm}-\hspace{-0.08truecm}(j\hspace{-0.08truecm}+\hspace{-0.08truecm}\frac{1}{2})\eta)
                           &\L_k(u\hspace{-0.08truecm}-\hspace{-0.08truecm}(j\hspace{-0.08truecm}+\hspace{-0.08truecm}\frac{1}{2})\eta)
                           &\hspace{-0.08truecm}-\hspace{-0.08truecm}a(u\hspace{-0.08truecm}-\hspace{-0.08truecm}(j\hspace{-0.08truecm}+\hspace{-0.08truecm}\frac{1}{2})\eta)\\[4pt]
                           &&&\hspace{-0.08truecm}-\hspace{-0.08truecm}d(u\hspace{-0.08truecm}-\hspace{-0.08truecm}(j\hspace{-0.08truecm}-\hspace{-0.08truecm}\frac{1}{2})\eta)
                           &\L_k(u\hspace{-0.08truecm}-\hspace{-0.08truecm}(j\hspace{-0.08truecm}-\hspace{-0.08truecm}\frac{1}{2})\eta)
                         \end{array}
\hspace{-0.12truecm}\rt|,\no\\[8pt]
&&\quad\quad j=\frac{1}{2},1,\frac{3}{2},\cdots,\label{Eigenvlue-2}
\eea} where the functions $a(u)$ and $d(u)$ are given by
(\ref{a-function}) and (\ref{d-function}). For example, the first
three ones are given by \bea
  \L_k^{(1)}(u)&=& \L_k(u+\frac{\eta}{2})\,\L_k(u-\frac{\eta}{2})-\d(u+\frac{\eta}{2}),\no\\[2pt]
  \L_k^{(\frac{3}{2})}(u)&=& \L_k(u+\eta)\,\L_k(u)\,\L_k(u-\eta)
                           -\d(u+\eta)\,\L_k(u-\eta)-\d(u)\,\L_k(u+\eta),\no\\[2pt]
  \L_k^{(2)}(u)&=& \L_k(u+\frac{3\eta}{2})\,\L_k(u+\frac{\eta}{2})
                   \,\L_k(u-\frac{\eta}{2})\,\L_k(u-\frac{3\eta}{2})\no\\[2pt]
              &&-\d(u+\frac{3\eta}{2})\,\L_k(u-\frac{\eta}{2})\,\L_k(u-\frac{3\eta}{2})-\d(u+\frac{\eta}{2})\,\L_k(u+\frac{3\eta}{2})\,\L_k(u-\frac{3\eta}{2})\no\\[2pt]
              &&-\d(u-\frac{\eta}{2})\,\L_k(u+\frac{3\eta}{2})\,\L_k(u+\frac{\eta}{2})+\d(u+\frac{3\eta}{2})\d(u-\frac{\eta}{2}).\no
\eea
The truncation identity (\ref{Truncation-oper}) of the spin-$\frac{p}{2}$ transfer matrix leads to the fact that the corresponding
eigenvalue $\L^{(\frac{p}{2})}_k(u)$ satisfies the relation
\bea
\L_k^{(\frac{p}{2})}(u)={\bf {\cal{A}}}(u)+ {\bf {\cal{D}}}(u)
       +\d(u-(\frac{p-1}{2})\eta)\L_k^{(\frac{p-2}{2})}(u),\label{Eigen-truncation}
\eea  where the functions ${\bf {\cal{A}}}(u)$ and ${\bf {\cal{D}}}(u)$  are given by (\ref{Average-1})-(\ref{Average-2}).

It is believed \cite{Baz90,Bax04,Geh06} that the quasi-periodicity (\ref{Eigen-periodic}), the asymptotic behavior (\ref{Eigen-Asymp}),
the analytic property (\ref{Eigenvalue-Anal}) and
the truncation identity (\ref{Eigen-truncation}) completely determine the eigenvalues $\{\L_k(u)|k=1,2,\cdots,p\}$
of the fundamental transfer matrix $t(u)$ given by (\ref{transfer-matrix}).

\subsection{T-Q relation}
\subsubsection{Generic case}

Following the method developed in \cite{cao1}  (or for details we
refer the reader to \cite{Wan15}), let us introduce the following
inhomogeneous $T-Q$ relation \bea
\L_k(u)=e^{\phi_k}a(u)\frac{Q(u-\eta)}{Q(u)}+e^{-\phi_k}d(u)\frac{Q(u+\eta)}{Q(u)}+2^{(1-p)N}c_k\,\frac{F_k(u)}{Q(u)},\label{T-Q-relation}
\eea where $\phi_k$ is a generic complex number \footnote{$\phi_k$ is chosen such that the degree of the trigonometric
polynomial $F_k(u)$ given by (\ref{F-fuction}) is $pN$.}, the
functions $a(u)$ and $d(u)$ are given by (\ref{a-function}) and
(\ref{d-function}), the function $F_k(u)$ is given by \bea
&&F_k(u)={\bf {\cal{A}}}(u)+ {\bf {\cal{D}}}(u)-e^{p\phi_k}{\bf \bar {\cal{A}}}(u)-e^{-p\phi_k}{\bf \bar {\cal{D}}}(u), \label{F-fuction}\\
&&{\bf \bar {\cal{A}}}(u)=\prod_{m=1}^p a(u-m\eta),\quad {\bf \bar {\cal{D}}}(u)=\prod_{m=1}^p d(u-m\eta),\label{Bar-a-d-functions}
\eea
and the function $Q(u)$ is a trigonometric polynomial of degree $(p-1)N$
\bea
Q(u)=\prod_{j=1}^{(p-1)N}\sinh(u-\l_j).\label{Q-function}
\eea Here the $(p-1)N+1$ parameters $c_k$ and $\{\l_j|j=1,\cdots,(p-1)N\}$ satisfy the associated Bethe Ansatz equations (BAEs)
\bea
 && e^{\phi_k}a(\l_j)Q(\l_j - \eta) +
     e^{ - \phi_k}d(\l_j)Q(\l_j + \eta) \no \\[6pt]
     && \qquad
      + 2^{(1-p)N}c_kF_k(\l_j)=0,\quad j=1,\cdots,(p-1)N,\label{BAE-1} \\[6pt]
 && q^k
     D^{(+)}+q^{-k}F^{(+)}-2\lt\{D^{(+)}F^{(+)}\rt\}^{\frac{1}{2}}\cosh(\phi_k+\frac{3}{2}N\eta)
    \no\\[6pt]
 &&  \quad\quad=c_k \,e^{\sum_{j=1}^{(p-1)N}\l_j}\lt\{\{D^{(+)}\}^p+\{F^{(+)}\}^p
-2(-1)^N\lt\{D^{(+)}F^{(+)}\rt\}^{\frac{p}{2}}\cosh p\phi_k\rt\},\label{BAE-2}\\[6pt]
 && q^{-k}D^{(-)}+q^{k}F^{(-)}-(-1)^Ne^{\phi_k-\frac{N}{2}\eta}\lt\{
\frac{G^{(-)}H^{(+)}}{\lt\{D^{(+)}F^{(+)}\rt\}^{\frac{1}{2}}}
+e^{-2\phi_k+N\eta}\frac{G^{(+)}H^{(-)}}{\lt\{D^{(+)}F^{(+)}\rt\}^{\frac{1}{2}}}\rt\}\no\\[6pt]
 && \quad\quad= c_ke^{ -
   \sum_{j=1}^{(p-1)N}\l_j} \no \\[6pt] &&\qquad \quad \times  \lt\{ \{D^{(-)}\}^p
    + \{F^{(-)}\}^p
    - e^{p\phi_k}\frac{\{G^{(-)}H^{(+)}\}^p}{\lt\{D^{(+)}F^{(+)}\rt\}^{\frac{p}{2}}}
    - e^{-p\phi_k}\frac{\{G^{(+)}H^{(-)}\}^p}{\lt\{D^{(+)}F^{(+)}\rt\}^{\frac{p}{2}}}
\rt\}.\label{BAE-3} \eea Here the  constants $D^{(\pm)}$ and
$F^{(\pm)}$ are given by (\ref{Constant-1}) and $G^{(\pm)}$ and
$H^{(\pm)}$ read \bea G^{(\pm)}=\prod_{n=1}^Ng_n^{(\pm)},\quad
H^{(\pm)}=\prod_{n=1}^Nh_n^{(\pm)}. \label{Constant-2} \eea
Notice that for a given $\phi_k$, either (\ref{BAE-2}) or (\ref{BAE-3}) only serves as a selection rule
(see the remarks in the end of appendix A).

It can be shown that the inhomogeneous $T-Q$ relation
(\ref{T-Q-relation}) does indeed satisfy
(\ref{Eigen-periodic})-(\ref{Eigenvalue-Anal}) and
(\ref{Eigen-truncation}) provided that the $(p-1)N+1$ parameters
$c_k$ and $\{\l_j|j=1,\cdots,(p-1)N\}$ obey the BAEs
(\ref{BAE-1})-(\ref{BAE-3}). The proof is given in appendix A. Hence
$\{\L_k(u)|k=1,\cdots,p\}$ given by the $T-Q$ relation
(\ref{T-Q-relation}) become the eigenvalues of the transfer matrix
$t(u)$ of the $\tau_2$-model with periodic boundary condition.

Numerical solutions of the Bethe Ansatz equations and exact
diagonalization of the transfer matrix are performed for $p=3$,
$N=2$ and $N=3$ and arbitrarily chosen inhomogeneity parameters. The
Bethe roots for given $\phi_k$ are shown in Table 1\&2 ($N=2$) and
Table 3\&4 ($N=3$) respectively. The $\L(u)$ curves calculated from
exact diagonalization and from the $T-Q$ relation coincide exactly
(Figure 1\&2), which imply that the inhomogeneous $T-Q$ relation
does indeed give the complete and correct spectrum of the generic
$\tau_2$ transfer matrix.

With the help of the determinant representation (\ref{Eigenvlue-2}),
we can obtain the eigenvalues
$\{\Lambda^{(j)}(u)|j=1,\frac{3}{2},\cdots,\frac{p}{2}\}$ of the
higher spin-$j$ transfer matrices
$\{t^{(j)}(u)|j=1,\frac{3}{2},\cdots,\frac{p}{2}\}$.

\begin{figure}[ht]
\begin{center}
  \includegraphics[width=10cm]{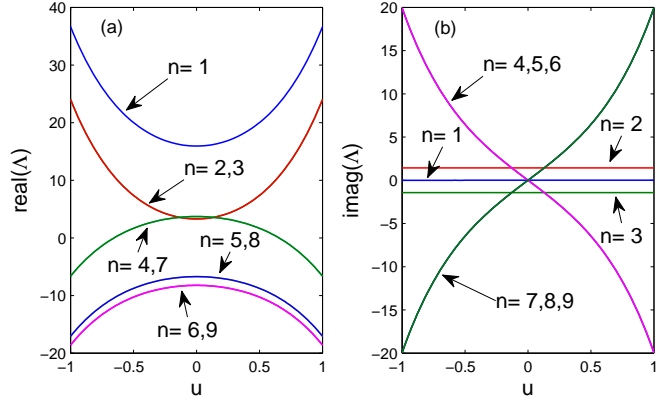}\\
\caption{Real (a) and imaginary (b) parts of the eigenvalues
$\Lambda(u)$ for $p=3$, $N=2$, $d_1^{(\pm)}=2$, $f_1^{(\pm)}=1/2$,
$g_1^{(\pm)}=3$, $h_1^{(\pm)}=1/3$, $d_2^{(\pm)}=\sqrt{3}$,
$f_2^{(\pm)}=1/\sqrt{3}$, $g_2^{(\pm)}=\sqrt{2}$ and
$h_2^{(\pm)}=1/\sqrt{2}$. The curves calculated from exact
diagonalization coincide with those derived from the inhomogeneous
$T-Q$ relation. }\label{N2Lambda}
\end{center}
\end{figure}
\begin{table}
\caption{The Bethe roots solved from the Bethe Ansatz equations
(\ref{BAE-1})-(\ref{BAE-3}) for $p=3$, $N=2$ and $\phi_k=0$ with the
inhomogeneity parameters $d_1^{(\pm)}=2$, $f_1^{(\pm)}=1/2$,
$g_1^{(\pm)}=3$, $h_1^{(\pm)}=1/3$, $d_2^{(\pm)}=\sqrt{3}$,
$f_2^{(\pm)}=1/\sqrt{3}$, $g_2^{(\pm)}=\sqrt{2}$ and
$h_2^{(\pm)}=1/\sqrt{2}$.} {\scriptsize
\begin{tabular}{ cc|cccc|c} \hline\hline $n$ & $k$ &
$\lambda_1$ & $\lambda_2$ & $\lambda_3$ & $\lambda_4$ &   $c_k$ \\
\hline
$1$ & $0$ & $-0.41481-0.48777i$ & $-0.41481+0.48777i$ & $0.16543-0.35104i$ & $0.16543+0.35104i$ & $0.07290+0.00000i$ \\
$2$ & $0$ & $-0.66826-1.49724i$ & $-0.04032-0.48519i$ & $0.06867+1.53930i$ & $0.14115+0.44313i$ & $0.07290+0.00000i$ \\
$3$ & $0$ & $-0.66826+1.49724i$ & $-0.04032+0.48519i$ & $0.06867-1.53930i$ & $0.14115-0.44313i$ & $0.07290+0.00000i$ \\
$4$ & $1$ & $-0.38066-1.28929i$ & $-0.18674+0.42053i$ & $0.33201+0.59239i$ & $0.48476+1.21602i$ & $-0.08872+0.02966i$ \\
$5$ & $1$ & $-0.21413-0.44969i$ & $0.07446+0.72464i$ & $0.16553-0.57279i$ & $0.22352+1.23748i$ & $-0.08872+0.02966i$ \\
$6$ & $1$ & $-0.22477-0.43023i$ & $0.06454+0.56084i$ & $0.18032-0.56761i$ & $0.22930+1.37666i$ & $-0.08872+0.02966i$ \\
$7$ & $2$ & $-0.38066+1.28929i$ & $-0.18674-0.42053i$ & $0.33201-0.59239i$ & $0.48476-1.21602i$ & $-0.08872-0.02966i$ \\
$8$ & $2$ & $-0.21413+0.44969i$ & $0.07446-0.72464i$ & $0.16553+0.57279i$ & $0.22352-1.23748i$ & $-0.08872-0.02966i$ \\
$9$ & $2$ & $-0.22477+0.43023i$ & $0.06454-0.56084i$ & $0.18032+0.56761i$ & $0.22930-1.37666i$ & $-0.08872-0.02966i$ \\
\hline\hline \end{tabular}}
\end{table}
\begin{table}
\caption{The Bethe roots solved from the Bethe Ansatz equations
(\ref{BAE-1})-(\ref{BAE-3}) for $p=3$, $N=2$ and $\phi_k=1$ with the
inhomogeneity parameters $d_1^{(\pm)}=2$, $f_1^{(\pm)}=1/2$,
$g_1^{(\pm)}=3$, $h_1^{(\pm)}=1/3$, $d_2^{(\pm)}=\sqrt{3}$,
$f_2^{(\pm)}=1/\sqrt{3}$, $g_2^{(\pm)}=\sqrt{2}$ and
$h_2^{(\pm)}=1/\sqrt{2}$.} {\scriptsize
\begin{tabular}{ cc|cccc|c} \hline\hline $n$ & $k$ &
$\lambda_1$ & $\lambda_2$ & $\lambda_3$ & $\lambda_4$ &   $c_k$ \\
\hline
$1$ & $0$ & $-1.11708+0.45660i$ & $-0.21978-0.33183i$ & $0.12531+0.21565i$ & $0.14087-0.52391i$ & $0.08911+0.01654i$ \\
$2$ & $0$ & $-1.24428+1.38621i$ & $-0.01821-0.44011i$ & $0.04372-1.51711i$ & $0.14809+0.38754i$ & $0.08911+0.01654i$ \\
$3$ & $0$ & $-1.32148+1.26038i$ & $-0.00660+0.42108i$ & $0.12388-0.38684i$ & $0.13352-1.47809i$ & $0.08911+0.01654i$ \\
$4$ & $1$ & $-0.79677-1.04629i$ & $-0.63157+0.66772i$ & $0.09800+0.34342i$ & $1.41424+1.03879i$ & $-0.21364+0.11605i$ \\
$5$ & $1$ & $-0.54458-0.56692i$ & $-0.39125+1.02018i$ & $0.18039-0.47610i$ & $0.83934+1.02649i$ & $-0.21364+0.11605i$ \\
$6$ & $1$ & $-0.50863-0.52803i$ & $-0.23822+1.02316i$ & $0.21632-0.51790i$ & $0.61444+1.02642i$ & $-0.21364+0.11605i$ \\
$7$ & $2$ & $-0.16368-0.27977i$ & $-0.15197-1.49634i$ & $0.56109+1.39587i$ & $0.74134-0.43992i$ & $-0.09374-0.03046i$ \\
$8$ & $2$ & $-0.17493+0.27850i$ & $0.03704-0.71000i$ & $0.53931+0.65255i$ & $0.58536-1.04121i$ & $-0.09374-0.03046i$ \\
$9$ & $2$ & $-0.18197+0.26803i$ & $0.06268-0.56937i$ & $0.53093+0.61846i$ & $0.57514-1.13728i$ & $-0.09374-0.03046i$ \\
\hline\hline \end{tabular}}
\end{table}
\begin{figure}[ht]
\begin{center}
  \includegraphics[width=10cm]{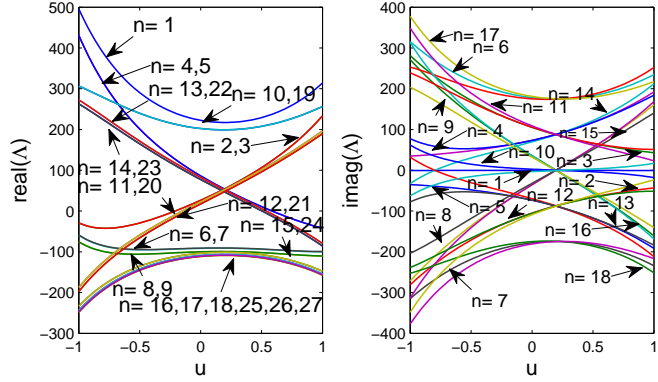}\\
\caption{Real (a) and imaginary (b) parts of the eigenvalues
$\Lambda(u)$ for $p=3$, $N=3$, $d_{1, 2, 3}^{(+)} = \{2, 0.2, 3\}$,
$f_{1, 2, 3}^{(-)}=\{0.6, 4, 0.5\}$, $g_{1, 2, 3}^{(-)}= \{1, 0.4,
5\}$, $h_{1, 2, 3}^{(-)}= \{1.2, 2, 0.3\}$, $d_{1, 2, 3}^{(-)}= \{3,
1, 1.5\}$, $f_{1, 2, 3}^{(+)}=\{0.4, 0.8, 1\}$, $g_{1, 2,
3}^{(+)}=\{4, 0.1, 2\}$ and $h_{1, 2, 3}^{(+)}=\{0.3, 8, 0.75\}$.
The curves calculated from exact diagonalization coincide with those
derived from the inhomogeneous $T-Q$ relation. }\label{N3Lambda}
\end{center}
\end{figure}
\begin{landscape}
\begin{table}
\caption{The Bethe roots solved from the Bethe Ansatz equations
(\ref{BAE-1})-(\ref{BAE-3}) for $p=3$, $N=3$ and $\phi_k=0$ with the
inhomogeneity parameters $d_{1, 2, 3}^{(+)} = \{2, 0.2, 3\}$, $f_{1,
2, 3}^{(-)}=\{0.6, 4, 0.5\}$, $g_{1, 2, 3}^{(-)}=\{1, 0.4, 5\}$,
$h_{1, 2, 3}^{(-)}= \{1.2, 2, 0.3\}$, $d_{1, 2, 3}^{(-)}= \{3, 1,
1.5\}$, $f_{1, 2, 3}^{(+)} = \{0.4, 0.8, 1\}$, $g_{1, 2, 3}^{(+)} =
\{4, 0.1, 2\}$ and $h_{1, 2, 3}^{(+)} = \{0.3, 8, 0.75\}$. }
{\scriptsize
\begin{tabular}{ cc| cccccc|c} \hline\hline $n$ & $k$ & $\lambda_1$
& $\lambda_2$ &
$\lambda_3$ & $\lambda_4$ & $\lambda_5$ & $\lambda_6$ &   $c_k$ \\
\hline
$1$ & $0$ &$-1.40457-0.54217i$ & $-1.14643+0.48574i$ & $0.21213-0.51244i$ & $0.26215+0.52628i$ & $2.23260+0.58946i$ & $2.25229-0.54688i$ & $0.11101+0.00000i$ \\
$2$ & $0$ &$-1.37944+0.44821i$ & $-1.08099-1.53477i$ & $0.15169-0.57167i$ & $0.25002-1.51348i$ & $2.22412+0.58246i$ & $2.24278-0.55234i$ & $-0.11101-0.00000i$ \\
$3$ & $0$ &$-1.34404+1.57012i$ & $-1.20296-0.43961i$ & $0.17315+0.54669i$ & $0.31772+1.41399i$ & $2.22188+0.59253i$ & $2.24243-0.54213i$ & $-0.11101-0.00000i$ \\
$4$ & $0$ &$-1.40652-0.53734i$ & $-1.16460+0.49706i$ & $0.16933-0.47789i$ & $0.21026+1.53395i$ & $2.25354+0.54696i$ & $2.34616-1.56275i$ & $0.11101-0.00000i$ \\
$5$ & $0$ &$-1.40978-0.54088i$ & $-1.16556+0.47933i$ & $0.19408+0.51898i$ & $0.27686-1.43479i$ & $2.21747-0.59576i$ & $2.29509-1.56846i$ & $-0.11101+0.00000i$ \\
$6$ & $0$ &$-1.34697+1.55928i$ & $-1.19166-0.45593i$ & $0.16445+0.47419i$ & $0.18455-0.54318i$ & $2.25285+0.53773i$ & $2.34494+1.56950i$ & $-0.11101-0.00000i$ \\
$7$ & $0$ &$-1.37026+0.45267i$ & $-1.08588-1.50001i$ & $0.16322+0.56858i$ & $0.18792-0.51432i$ & $2.21829-0.58781i$ & $2.29488-1.56071i$ & $-0.11101-0.00000i$ \\
$8$ & $0$ &$-1.37070+0.44175i$ & $-1.04295-1.51148i$ & $-0.00426-1.54904i$ & $0.21005+0.50411i$ & $2.26182+0.54164i$ & $2.35421-1.56857i$ & $-0.11101-0.00000i$ \\
$9$ & $0$ &$-1.33477+1.56165i$ & $-1.18061-0.43623i$ & $0.13757+1.54896i$ & $0.25827-0.51762i$ & $2.22542-0.59242i$ & $2.30229-1.56434i$ & $0.11101-0.00000i$ \\
$10$ & $1$ &$-2.36717-0.64080i$ & $-1.31215+0.72735i$ & $0.21062-0.51316i$ & $0.25655+0.52768i$ & $1.95573-1.23307i$ & $1.97213+0.77216i$ & $0.15671-0.11901i$ \\
$11$ & $1$ &$-2.25583+1.55193i$ & $-1.41980-0.27051i$ & $0.17387+0.54910i$ & $0.31057+1.39847i$ & $1.94548-1.22217i$ & $1.96142+0.77494i$ & $-0.15671+0.11901i$ \\
$12$ & $1$ &$-2.26897+0.33840i$ & $-1.32210-1.28008i$ & $0.15296-0.57835i$ & $0.24772-1.50723i$ & $1.94115-1.23792i$ & $1.96495+0.76376i$ & $-0.15671+0.11901i$ \\
$13$ & $1$ &$-2.38318-0.63525i$ & $-1.32403+0.71167i$ & $0.19433+0.51841i$ & $0.27035-1.42332i$ & $1.94420+0.71893i$ & $2.01404-0.25029i$ & $0.15671-0.11901i$ \\
$14$ & $1$ &$-2.37051-0.62354i$ & $-1.33281+0.72828i$ & $0.17173-0.47490i$ & $0.20767+1.53107i$ & $1.97436-1.27414i$ & $2.06528-0.24660i$ & $0.15671-0.11901i$ \\
$15$ & $1$ &$-2.24810+0.32841i$ & $-1.31046-1.25036i$ & $0.01507-1.55164i$ & $0.20876+0.50544i$ & $1.98173-1.28189i$ & $2.06872-0.25140i$ & $-0.15671+0.11901i$ \\
$16$ & $1$ &$-2.24610+1.53155i$ & $-1.40132-0.26168i$ & $0.14240+1.55026i$ & $0.25296-0.51903i$ & $1.94771+0.72751i$ & $2.02007-0.24686i$ & $-0.15671+0.11901i$ \\
$17$ & $1$ &$-2.26756+1.53313i$ & $-1.40348-0.28154i$ & $0.16668+0.47080i$ & $0.18558-0.54463i$ & $1.97161-1.28526i$ & $2.06290-0.25235i$ & $0.15671-0.11901i$ \\
$18$ & $1$ &$-2.24957+0.35238i$ & $-1.34102-1.25755i$ & $0.16508+0.57440i$ & $0.18837-0.51321i$ & $1.93836+0.72705i$ & $2.01449-0.24291i$ & $0.15671-0.11901i$ \\
$19$ & $2$ &$-1.96565-0.85298i$ & $-1.71352+0.65345i$ & $0.21076-0.51328i$ & $0.25633+0.52691i$ & $1.95534-0.73070i$ & $1.97245+1.27644i$ & $0.15671+0.11901i$ \\
$20$ & $2$ &$-1.90791+0.17785i$ & $-1.68438-1.40896i$ & $0.15356-0.57754i$ & $0.24708-1.50880i$ & $1.94392-0.73195i$ & $1.96344+1.26765i$ & $-0.15671-0.11901i$ \\
$21$ & $2$ &$-1.93248+1.31745i$ & $-1.74683-0.32459i$ & $0.17394+0.54847i$ & $0.31135+1.40051i$ & $1.94926-0.72122i$ & $1.96047+1.28081i$ & $-0.15671-0.11901i$ \\
$22$ & $2$ &$-1.97499-0.84383i$ & $-1.72780+0.65878i$ & $0.17166-0.47433i$ & $0.20771+1.53004i$ & $1.97521-0.77013i$ & $2.06392+0.25931i$ & $0.15671+0.11901i$ \\
$23$ & $2$ &$-1.97821-0.85630i$ & $-1.72547+0.64443i$ & $0.19443+0.51789i$ & $0.26961-1.42148i$ & $1.94102+1.22018i$ & $2.01433+0.25512i$ & $0.15671+0.11901i$ \\
$24$ & $2$ &$-1.91801+1.30696i$ & $-1.72979-0.32555i$ & $0.14265+1.54942i$ & $0.25294-0.51873i$ & $1.94809+1.23029i$ & $2.01983+0.25904i$ & $-0.15671-0.11901i$ \\
$25$ & $2$ &$-1.89125+0.17854i$ & $-1.66717-1.39606i$ & $0.01519-1.54823i$ & $0.20867+0.50484i$ & $1.98138-0.77737i$ & $2.06890+0.25654i$ & $-0.15671-0.11901i$ \\
$26$ & $2$ &$-1.93408+1.29994i$ & $-1.73940-0.33954i$ & $0.16686+0.47060i$ & $0.18568-0.54466i$ & $1.97229-0.77985i$ & $2.06435+0.25334i$ & $0.15671+0.11901i$ \\
$27$ & $2$ &$-1.90020+0.19277i$ & $-1.68723-1.38868i$ & $0.16492+0.57395i$ & $0.18835-0.51320i$ & $1.93588+1.23266i$ & $2.01399+0.26233i$ & $0.15671+0.11901i$ \\
\hline\hline \end{tabular}}
\end{table}
\end{landscape}
\begin{landscape}
\begin{table}
\caption{The Bethe roots solved from the Bethe Ansatz equations
(\ref{BAE-1})-(\ref{BAE-3}) for  $p=3$, $N=3$ and $\phi_k=1$ with the
inhomogeneity parameters $d_{1, 2, 3}^{(+)} = \{2, 0.2, 3\}$, $f_{1,
2, 3}^{(-)}=\{0.6, 4, 0.5\}$, $g_{1, 2, 3}^{(-)}= \{1, 0.4, 5\}$,
$h_{1, 2, 3}^{(-)}= \{1.2, 2, 0.3\}$, $d_{1, 2, 3}^{(-)}= \{3, 1,
1.5\}$, $f_{1, 2, 3}^{(+)} = \{0.4, 0.8, 1\}$, $g_{1, 2, 3}^{(+)} =
\{4, 0.1, 2\}$ and $h_{1, 2, 3}^{(+)} = \{0.3, 8, 0.75\}$. }
{\scriptsize
\begin{tabular}{ cc| cccccc|c} \hline\hline $n$ & $k$ & $\lambda_1$
& $\lambda_2$ & $\lambda_3$ & $\lambda_4$ & $\lambda_5$ &
$\lambda_6$ &   $c_k$ \\ \hline
$1$ & $0$ &$-1.18077-0.63093i$ & $-0.77548+0.42106i$ & $0.20506-0.48775i$ & $0.41948+0.54567i$ & $1.84319-0.63232i$ & $2.12005+0.78427i$ & $0.03770-0.00000i$ \\
$2$ & $0$ &$-1.10529+0.34161i$ & $-0.62163-1.46040i$ & $0.05537-0.62831i$ & $0.37145-1.52447i$ & $1.82712-0.62376i$ & $2.10450+0.75374i$ & $-0.03770-0.00000i$ \\
$3$ & $0$ &$-1.09702-1.53557i$ & $-0.90813-0.34431i$ & $0.08547+0.52810i$ & $0.62553+1.18619i$ & $1.84220-0.61528i$ & $2.08348+0.78086i$ & $0.03770+0.00000i$ \\
$4$ & $0$ &$-1.18259-0.61106i$ & $-0.80832+0.44954i$ & $0.12775-0.40794i$ & $0.16347-1.54513i$ & $1.92326+0.29737i$ & $2.40796-1.32436i$ & $-0.03770+0.00000i$ \\
$5$ & $0$ &$-1.19765-0.62056i$ & $-0.80874+0.41520i$ & $0.27139+0.55866i$ & $0.41527-1.22883i$ & $1.66113+1.39627i$ & $2.29013-0.52074i$ & $0.03770+0.00000i$ \\
$6$ & $0$ &$-1.10378-1.56716i$ & $-0.87367-0.36252i$ & $0.10949+0.38463i$ & $0.16998-0.54536i$ & $1.91951+0.28868i$ & $2.41000-1.33987i$ & $-0.03770-0.00000i$ \\
$7$ & $0$ &$-1.07093+0.34409i$ & $-0.63932-1.35059i$ & $0.16445+0.63300i$ & $0.22628-0.53500i$ & $1.67027+1.41723i$ & $2.28078-0.50872i$ & $0.03770+0.00000i$ \\
$8$ & $0$ &$-1.08780+0.31120i$ & $-0.46010-1.19085i$ & $-0.36385+1.47511i$ & $0.19124+0.44571i$ & $1.92905+0.28943i$ & $2.42300-1.33060i$ & $0.03770-0.00000i$ \\
$9$ & $0$ &$-1.06678-1.55890i$ & $-0.87270-0.31728i$ & $0.22250+1.47739i$ & $0.37070-0.49410i$ & $1.68121+1.39916i$ & $2.29661-0.50628i$ & $0.03770+0.00000i$ \\
$10$ & $1$ &$-1.45639-0.68240i$ & $-0.79897+0.53967i$ & $0.20441-0.49025i$ & $0.40545+0.55500i$ & $1.67944-0.96068i$ & $1.83770+0.82124i$ & $0.03029-0.01164i$ \\
$11$ & $1$ &$-1.33685-1.52407i$ & $-0.96823-0.26543i$ & $0.07943+0.53134i$ & $0.60852+1.14853i$ & $1.67359-0.92257i$ & $1.81519+0.81477i$ & $0.03029-0.01164i$ \\
$12$ & $1$ &$-1.32102+0.27890i$ & $-0.69844-1.31874i$ & $0.04696-0.65235i$ & $0.37397-1.51493i$ & $1.63642-0.95097i$ & $1.83377+0.79908i$ & $-0.03029+0.01164i$ \\
$13$ & $1$ &$-1.48351-0.67132i$ & $-0.83065+0.52522i$ & $0.27356+0.56560i$ & $0.39269-1.21484i$ & $1.54772+0.85431i$ & $1.97184-0.27640i$ & $0.03029-0.01164i$ \\
$14$ & $1$ &$-1.46147-0.65184i$ & $-0.83885+0.56232i$ & $0.13313-0.40377i$ & $0.16379-1.54201i$ & $1.87621-0.08166i$ & $1.99885-1.24207i$ & $-0.03029+0.01164i$ \\
$15$ & $1$ &$-1.28987+0.24571i$ & $-0.62143-1.14415i$ & $-0.29341+1.56827i$ & $0.19063+0.44950i$ & $1.88066-0.09019i$ & $2.00508-1.24657i$ & $0.03029-0.01164i$ \\
$16$ & $1$ &$-1.30182-1.56123i$ & $-0.93594-0.24033i$ & $0.22228+1.48496i$ & $0.35819-0.50515i$ & $1.55627+0.87479i$ & $1.97267-0.27048i$ & $0.03029-0.01164i$ \\
$17$ & $1$ &$-1.34976-1.56888i$ & $-0.93296-0.28003i$ & $0.11438+0.37805i$ & $0.17059-0.54750i$ & $1.87114-0.09041i$ & $1.99826-1.25026i$ & $-0.03029+0.01164i$ \\
$18$ & $1$ &$-1.27444+0.29316i$ & $-0.74475-1.23106i$ & $0.16793+0.65364i$ & $0.22647-0.53682i$ & $1.52925+0.87589i$ & $1.96718-0.27224i$ & $0.03029-0.01164i$ \\
$19$ & $2$ &$-1.35404-0.79738i$ & $-0.89775+0.51195i$ & $0.20503-0.48909i$ & $0.40262+0.55358i$ & $1.63222-0.72436i$ & $1.88358+1.16274i$ & $0.03029+0.01164i$ \\
$20$ & $2$ &$-1.23264+0.20519i$ & $-0.79546-1.39083i$ & $0.05445-0.64797i$ & $0.36607-1.52397i$ & $1.59919-0.70425i$ & $1.88004+1.13766i$ & $-0.03029-0.01164i$ \\
$21$ & $2$ &$-1.28301+1.50079i$ & $-1.03057-0.29809i$ & $0.08365+0.52808i$ & $0.60821+1.16322i$ & $1.63587-0.68972i$ & $1.85750+1.15474i$ & $-0.03029-0.01164i$ \\
$22$ & $2$ &$-1.36454-0.77340i$ & $-0.93180+0.53133i$ & $0.13333-0.40114i$ & $0.16253-1.54377i$ & $1.83391+0.10660i$ & $2.03823-0.84378i$ & $-0.03029-0.01164i$ \\
$23$ & $2$ &$-1.37931-0.79345i$ & $-0.92608+0.49850i$ & $0.27348+0.56123i$ & $0.38926-1.19929i$ & $1.42181+1.05767i$ & $2.09248+0.09277i$ & $0.03029+0.01164i$ \\
$24$ & $2$ &$-1.24340+1.47355i$ & $-0.99698-0.27755i$ & $0.22686+1.48133i$ & $0.35590-0.50156i$ & $1.44032+1.08390i$ & $2.08896+0.09936i$ & $-0.03029-0.01164i$ \\
$25$ & $2$ &$-1.20219+0.18087i$ & $-0.67017-1.27621i$ & $-0.33153-1.52955i$ & $0.18985+0.44636i$ & $1.84149+0.09855i$ & $2.04421-0.84418i$ & $-0.03029-0.01164i$ \\
$26$ & $2$ &$-1.28768+1.45716i$ & $-0.99898-0.31718i$ & $0.11502+0.37783i$ & $0.17053-0.54662i$ & $1.83151+0.09548i$ & $2.04125-0.84925i$ & $0.03029+0.01164i$ \\
$27$ & $2$ &$-1.19567+0.22143i$ & $-0.81573-1.31025i$ & $0.16976+0.64979i$ & $0.22613-0.53479i$ & $1.40246+1.09819i$ & $2.08470+0.09306i$ & $0.03029+0.01164i$ \\
\hline\hline \end{tabular}}
\end{table}
\end{landscape}

\subsubsection{Degenerate case}
For generic inhomogeneous parameters
$\{d^{(\pm)}_n,~f^{(\pm)}_n,~g^{(\pm)}_n,~h^{(\pm)}_n
|n=1,\cdots,N\}$ obeying the constraint
(\ref{Constraint-1}), the inhomogeneous term
in the $T-Q$ relation (\ref{T-Q-relation}) does not  vanish. In this
subsection we consider some special case such that the inhomogeneous
term vanishes. In this case,  the inhomogeneous parameters
$\{d^{(\pm)}_n,~f^{(\pm)}_n,~g^{(\pm)}_n,~h^{(\pm)}_n
|n=1,\cdots,N\}$ have to obey some further constraints besides
(\ref{Constraint-1}) as follows: \bea
e^{2M\eta}\,G^{(-)}\,H^{(+)}=(-1)^NF^{(+)}\,F^{(-)},\label{Constraint-2}
\eea or \bea
e^{2M\eta}\,G^{(-)}\,H^{(+)}=(-1)^ND^{(+)}\,D^{(-)},\label{Constraint-3}
\eea   and \bea
F_k^{(p(N-2l))}(\{d^{(\pm)}_n,~f^{(\pm)}_n,~g^{(\pm)}_n,~h^{(\pm)}_n\})=0,\quad
l=1,\cdots,N-1.\label{Constraint-4} \eea Here  $D^{(\pm)}$,
$F^{(\pm)}$, $G^{(-)}$ and $H^{(+)}$ are given by (\ref{Constant-1})
and (\ref{Constant-2}), and each $F_k^{(p(N-2l))}$ (given in
(\ref{F-function-expansion-1}) below) is a polynomial of the
inhomogeneous parameters
$\{d^{(\pm)}_n,~f^{(\pm)}_n,~g^{(\pm)}_n,~h^{(\pm)}_n
|n=1,\cdots,N\}$. It is noted that in (\ref{Constraint-2}) (or
(\ref{Constraint-3})), $M$ is some non-negative integer. The
corresponding inhomogeneous $T-Q$ relation (\ref{T-Q-relation}) then
reduces to the conventional one \cite{Bax82}: \bea
\L_k(u)=e^{-(\frac{N}{2}-M+k)\eta}a(u)\lt\{\frac{D^{(+)}}{F^{(+)}}\rt\}^{\frac{1}{2}}\frac{\bar
Q(u-\eta)}{\bar Q(u)}
+e^{(\frac{N}{2}-M+k)\eta}d(u)\lt\{\frac{F^{(+)}}{D^{(+)}}\rt\}^{\frac{1}{2}}\frac{\bar
Q(u+\eta)}{\bar Q(u)},\label{T-Q relation-2} \eea or \bea
\L_k(u)=e^{-(\frac{N}{2}-M+k)\eta}a(u)\lt\{\frac{F^{(+)}}{D^{(+)}}\rt\}^{\frac{1}{2}}\frac{\bar
Q(u-\eta)}{\bar Q(u)}
+e^{(\frac{N}{2}-M+k)\eta}d(u)\lt\{\frac{D^{(+)}}{F^{(+)}}\rt\}^{\frac{1}{2}}\frac{\bar
Q(u+\eta)}{\bar Q(u)},\label{T-Q relation-3} \eea where the function
$\bar Q(u)$ becomes \cite{Nep04,CLSW,Yan06,Fra07,Cao14,cao1,Wan15} \bea
\bar Q(u)=\prod_{j=1}^{M}\sinh(u-\l_j).\label{Q-function-1} \eea The
$M$ parameters $\{\l_j|j=1,\cdots,M\}$ satisfy the associated
BAEs \bea
 e^{-(N-2M+2k)\eta}\,\frac{D^{(+)}\,a(\l_j)}{F^{(+)}\,d(\l_j)}=-\frac{\bar Q(\l_j+\eta)}{\bar Q(\l_j-\eta)},\quad j=1,\cdots,M,\label{BAE-4}
\eea
or
\bea
 e^{-(N-2M+2k)\eta}\,\frac{F^{(+)}\,a(\l_j)}{D^{(+)}\,d(\l_j)}=-\frac{\bar Q(\l_j+\eta)}{\bar Q(\l_j-\eta)},\quad j=1,\cdots,M.\label{BAE-5}
\eea
The proof is given in Appendix B.

\section{Conclusions} \setcounter{equation}{0}

The most general cyclic representations of the quantum $\tau_2$-model (also known as Baxter-Bazhanov-Stroganov (BBS) model) with
periodic boundary condition   has been studied via the off-diagonal Bethe Ansatz method \cite{Wan15}. Based on the
the truncation identity (\ref{Truncation-oper}) of the fused transfer matrices, we construct the inhomogeneous $T-Q$ relation
(\ref{T-Q-relation}) of the eigenvalue of the fundamental transfer matrix $t(u)$ and the associated BAEs (\ref{BAE-1})-(\ref{BAE-3}).

It should be noted that for generic inhomogeneity parameters
$\{d^{(\pm)}_n, f^{(\pm)}_n, g^{(\pm)}_n, h^{(\pm)}_n $ $|n=1,
\cdots, N\}$  obeying the constraint (\ref{Constraint-1}),
the inhomogeneous term (i.e., the third term) in the $T-Q$ relation
(\ref{T-Q-relation}) {\it does not}  vanish, as long as one takes a polynomial $Q$ function. However, if these
inhomogeneity parameters  satisfy  the further constraints
(\ref{Constraint-2}) and (\ref{Constraint-4}) (or
(\ref{Constraint-3}) and (\ref{Constraint-4})), the corresponding
$T-Q$ relation reduces to the conventional one (\ref{T-Q
relation-2}) (or (\ref{T-Q relation-3})).

\section*{Acknowledgments}

The financial support from  the National Natural
Science Foundation of China (Grant Nos. 11174335, 11375141,
11374334, 11434013 and 11425522), the National Program for Basic Research of MOST
(973 project under Grant No. 2011CB921700), BCMIIS and the Strategic Priority Research Program
of CAS are gratefully. One of the authors X. Xu was also partially supported by the NWU graduate student
innovation fund No. YZZ14102.

Y. Wang acknowledges J.\,H.\,H. Perk and H. Au-Yang for helpful discussions in the chiral Potts model
and for drawing his attention to the references \cite{alc,p}.


\section*{Appendix A: Proof of the $T-Q$ relation}
\setcounter{equation}{0}
\renewcommand{\theequation}{A.\arabic{equation}}

In this appendix we prove that the inhomogeneous $T-Q$ relation (\ref{T-Q-relation}) does  satisfy (\ref{Eigen-periodic})-(\ref{Eigenvalue-Anal}) and (\ref{Eigen-truncation}) if the $(p-1)N+1$ parameters $c_k$ and $\{\l_j|j=1,\cdots,(p-1)N\}$ obey the BAEs (\ref{BAE-1})-(\ref{BAE-3}).

From the construction (\ref{T-Q-relation}) of the $T-Q$ relation  and the definitions (\ref{F-fuction})-(\ref{Q-function}),
one can easily check that the $T-Q$ relation satisfies the quasi-periodicity property (\ref{Eigen-periodic}). The BAEs
(\ref{BAE-2})-(\ref{BAE-3}) ensure that the asymptotic behavior (\ref{Eigen-Asymp}) is also fulfilled. Moreover the BAEs (\ref{BAE-1}) imply that
the functions given by the $T-Q$ relation (\ref{T-Q-relation}) actually satisfy (\ref{Eigenvalue-Anal}). So far, the $T-Q$ relation
already  makes (\ref{Eigen-periodic})-(\ref{Eigenvalue-Anal}) satisfied.

Let us consider the function $F_k(u)$ given by (\ref{F-fuction}).
For generic inhomogeneity parameters $\{d^{(\pm)}_n,~f^{(\pm)}_n,~g^{(\pm)}_n,~h^{(\pm)}_n |n=1,\cdots,N\}$ satisfying the constraint
(\ref{Constraint-1}), we know that, as a function
of $e^u$, $F_k(u)$ is a Laurent polynomial of degree $pN$ with the form
\bea
F_k(u)&=&{\bf {\cal{A}}}(u)+ {\bf {\cal{D}}}(u)-e^{p\phi_k}{\bf \bar {\cal{A}}}(u)-e^{-p\phi_k}{\bf \bar {\cal{D}}}(u)\no\\
&=&F_k^{(pN)}e^{pNu}+F_k^{(p(N-2))}e^{p(N-2)u}+\cdots+F_k^{(-pN)}e^{-pNu}\no\\
&=&{\cal{F}}_k^{(0)}\prod_{j=1}^{pN}\lt\{\frac{e^{u}}{e^{z_j}}-\frac{e^{z_j}}{e^u}\rt\}, \label{F-function-expansion}
\eea
where $\{z_j\,\,{\rm mod}\,(2i\pi)|j=1,\cdots,pN\}$ are the zeros of $F_k(u)$ which are all different from each other
and the constant ${\cal{F}}^{(0)}_k$ is related to the
asymptotic behaviors of the function. The relations (\ref{P-periodic}) and the definition (\ref{Bar-a-d-functions}) imply that
the function $F_k(u)$ actually is a Laurent polynomial of $e^{pu}$ with a degree $N$ (i.e., there are only $N+1$ non-vanishing
coefficients in the expansion (\ref{F-function-expansion}) ), namely,
\bea
 F_k(u)&=&\sum_{l=0}^NF_k^{(p(N-2l))}(\{d^{(\pm)}_n,~f^{(\pm)}_n,~g^{(\pm)}_n,~h^{(\pm)}_n\})e^{p(N-2l)u}\no\\
 &\stackrel{{\rm def}}{=}&\sum_{l=0}^NF_k^{(p(N-2l))}e^{p(N-2l)u},\label{F-function-expansion-1}
\eea where the $N+1$ non-vanishing coefficients $\{F_k^{(p(N-2l))}|l=0,1,\cdots,N\}$ are  polynomials of the inhomogeneity
parameters $\{d^{(\pm)}_n,~f^{(\pm)}_n,~g^{(\pm)}_n,~h^{(\pm)}_n |n=1,\cdots,N\}$. Moreover, it follows that
\bea
F_k(z_j)=F_k(z_j+m\eta)=0,\quad m\in \Zb.\label{zeros}
\eea

Let us introduce the function $g(u)$ which is given by
\bea
g(u)=
\L_k^{(\frac{p}{2})}(u)-{\bf {\cal{A}}}(u)- {\bf {\cal{D}}}(u)
       -\d(u-(\frac{p-1}{2})\eta)\L_k^{(\frac{p-2}{2})}(u),\label{g-function}
\eea where $\L_k^{(\frac{p}{2})}(u)$ and $\L_k^{(\frac{p-2}{2})}(u)$ are given by the determinant
representation (\ref{Eigenvlue-2}) with $\L_k(u)$ given by  (\ref{T-Q-relation}). From the above definition, one knows that
the function $g(u)$ as a function of $e^u$  is a Laurent polynomial of degree $pN$ of similar form  as (\ref{F-function-expansion}). Hence $g(u)$
is uniquely determined by its $pN+1$ points values such as $+\infty$ (or $-\infty$) and $\{z_j|j=1,\cdots,pN\}$. Thanks to the property (\ref{zeros}), we have
\bea
\hspace{-1.2truecm}\L_k(z_j+m\eta)&=&e^{\phi_k}\,a(z_j+m\eta)\,\frac{Q(z_j+m\eta-\eta)}{Q(z_j+m\eta)}+e^{-\phi_k}\,d(z_j+m\eta)\,
  \frac{Q(z_j+m\eta+\eta)}{Q(z_j+m\eta)},\\[6pt]
\hspace{-1.2truecm}&&\quad\quad m\in \Zb,\quad j=1,\cdots,pN.\no
\eea
Substituting the above relations into (\ref{Eigenvlue-2}) and noting the fact $p\eta=2i\pi$, after some tedious calculation, we
have
\bea
\L_k^{(\frac{p}{2})}(z_j)&=&e^{p\phi_k}\bar{{\bf {\cal{A}}}}(z_j)+e^{-p\phi_k} \bar{{\bf {\cal{D}}}}(z_j)
       +\d(z_j-(\frac{p-1}{2})\eta)\L_k^{(\frac{p-2}{2})}(z_j)\no\\[6pt]
&=&{\bf {\cal{A}}}(z_j)+ {\bf {\cal{D}}}(z_j)
       +\d(z_j-(\frac{p-1}{2})\eta)\L_k^{(\frac{p-2}{2})}(z_j),\quad j=1,\cdots,pN.\label{A-1}
\eea In deriving the second equality, we have used the fact: $F_k(z_j)=0$. Then (\ref{A-1}) implies that the function $g(u)$ vanishes at the points $z_j$, namely,
\bea
g(z_j)=0,\quad j=1,\cdots,pN.\label{A-2}
\eea The BAEs (\ref{BAE-2})-(\ref{BAE-3}) imply that the functions given by the $T-Q$ relation (\ref{T-Q-relation}) also satisfy
 (\ref{Eigen-Asymp}), which give rise to
 \bea
 \lim_{u\rightarrow \pm\infty}g(u)=0. \label{A-3}
 \eea (\ref{A-2})-(\ref{A-3}) imply that  $g(u)=0$. Namely, the inhomogeneous $T-Q$ relation (\ref{T-Q-relation}) does
satisfy (\ref{Eigen-truncation}). Therefore we can conclude that  $\{\L_k(u)|k=1,\cdots,p\}$ given by the $T-Q$ relation (\ref{T-Q-relation})
are the eigenvalues of the transfer matrix $t(u)$ of the $\tau_2$-model with periodic boundary condition provided that
the $(p-1)N+1$ parameters $c_k$ and $\{\l_j|j=1,\cdots,(p-1)N\}$ satisfy  the BAEs (\ref{BAE-1})-(\ref{BAE-3}).

Some remarks are in order. Due to the fact that $g(u)$ given by
(\ref{g-function}) as a function of $e^u$ is a Laurent polynomial of
degree $pN$, the relations \bea g(u)=0,\quad {\rm when }\,\, u=z_1,
\cdots, z_{pN}, +\infty,\no \eea are already sufficient to ensure
$g(u)=0$. This implies that the BAEs (\ref{BAE-1})-(\ref{BAE-2}) are
sufficient to guarantee that the $T-Q$ relation (\ref{T-Q-relation})
satisfy (\ref{Eigen-periodic}), (\ref{Eigenvalue-Anal}),
(\ref{Eigen-truncation}) and (\ref{Eigen-Asymp}) with the
$u\rightarrow +\infty$ limit. Then the BAE (\ref{BAE-3}) only plays
a role of the selection rule such that the $u\rightarrow -\infty$
behavior also matches.


\section*{Appendix B: Proof of the degenerate case}
\setcounter{equation}{0}
\renewcommand{\theequation}{B.\arabic{equation}}
In this appendix we show that the inhomogeneous $T-Q$ relation (\ref{T-Q-relation}) does  reduce to the conventional one (\ref{T-Q relation-2}) (or
(\ref{T-Q relation-3})) when the inhomogeneity parameters satisfy the constraints (\ref{Constraint-2}) and (\ref{Constraint-4}) or  (\ref{Constraint-3}) and (\ref{Constraint-4}).

Suppose that the inhomogeneous $T-Q$ relation (\ref{T-Q-relation})
can be reduced to the conventional one, namely, \bea
\L_k(u)=e^{\phi_k}a(u)\frac{\bar Q(u-\eta)}{\bar
Q(u)}+e^{-\phi_k}d(u)\frac{\bar Q(u+\eta)}{\bar Q(u)},
\label{T-Q-relation-B} \eea where  the $Q$-function is \bea \bar
Q(u)=\prod_{j=1}^{M}\sinh(u-\l_j),\no \eea and $M$ is a
non-negative integer to be specified by (\ref{Constraint-2}) (or
(\ref{Constraint-3})). The asymptotic behavior  (\ref{Eigen-Asymp})
of $\L_k(u)$ now becomes \bea &&q^k
      D^{(+)} + q^{-k}F^{(+)}
      - 2\lt\{D^{(+)}F^{(+)}\rt\}^{\frac{1}{2}}
     \cosh(\phi_k + (\frac{N}{2} - M)\eta)=0,  \\
 &&
      q^{-k}D^{(-)} + q^{k}F^{(-)}
      - (-1)^N \no \\ && \qquad \times \lt\{e^{\phi_k +
      (\frac{N}{2} + M)\eta}
      \frac{G^{(-)}H^{(+)}}{\lt\{D^{(+)}F^{(+)}\rt\}^{\frac{1}{2}}}
       +
      e^{-\phi_k - (\frac{N}{2} + M)\eta}
      \frac{G^{(+)}H^{(-)}}{\lt\{D^{(+)}F^{(+)}\rt\}^{\frac{1}{2}}}\rt\}=0.
\eea
Only when the inhomogeneity parameters obey the constraint (\ref{Constraint-2}) or (\ref{Constraint-3}), there does exist a solution to
the above two  equations:
\bea
\lt\{\begin{array}{ll}
e^{\phi_k}=e^{-(\frac{N}{2}-M+k)\eta}\lt\{\frac{D^{(+)}}{F^{(+)}}\rt\}^{\frac{1}{2}},&{\rm under}\,\,{\rm constraint}\, (\ref{Constraint-2}),\\
e^{\phi_k}=e^{-(\frac{N}{2}-M+k)\eta}\lt\{\frac{F^{(+)}}{D^{(+)}}\rt\}^{\frac{1}{2}},&{\rm under}\,\,{\rm constraint}\, (\ref{Constraint-3}).
\end{array}\rt.\label{B-1}
\eea
It is easy to check that both solutions give rise to $F_k^{(pN)}=F^{(-pN)}_k=0$. Together with (\ref{Constraint-4}), we have that in each constrained
case ((\ref{Constraint-2}) and (\ref{Constraint-4}) or (\ref{Constraint-3}) and (\ref{Constraint-4}))  the function $F_k(u)$ indeed vanishes, namely, $ F_k(u)=0$. Substituting the solution (\ref{B-1} ) into (\ref{T-Q-relation-B}), we obtain the conventional $T-Q$ relation (\ref{T-Q relation-2}) or (\ref{T-Q relation-3}) respectively.

Substituting (\ref{T-Q relation-2}) into (\ref{Eigenvlue-2}) and noting the fact $p\eta=2i\pi$, after some tedious calculation, we have
\bea
\L_k^{(\frac{p}{2})}(u)&=&e^{p\phi_k}\bar{{\bf {\cal{A}}}}(u)+e^{-p\phi_k} \bar{{\bf {\cal{D}}}}(u)
       +\d(u-(\frac{p-1}{2})\eta)\L_k^{(\frac{p-2}{2})}(u)\no\\
&=&{\bf {\cal{A}}}(u)+ {\bf {\cal{D}}}(u)
       +\d(u-(\frac{p-1}{2})\eta)\L_k^{(\frac{p-2}{2})}(u).\label{B-2}
\eea
In deriving the second equality, we have used the fact: $F_k(u)=0$ when  the inhomogeneity parameters
$\{d^{(\pm)}_n,~f^{(\pm)}_n,~g^{(\pm)}_n,~h^{(\pm)}_n |n=1,\cdots,N\}$  satisfy the constraints (\ref{Constraint-1}), (\ref{Constraint-2}) and (\ref{Constraint-4}). Similarly we can prove that  the reduced $T-Q$ relation (\ref{T-Q relation-3}) satisfies (\ref{Eigen-Asymp}) and
(\ref{Eigen-truncation})  provided that the inhomogeneity parameters obey the constraints (\ref{Constraint-1}), (\ref{Constraint-3}) and (\ref{Constraint-4}).



\begin{thebibliography}{99}

\bibitem{Baz90} V.\,V. Bazhanov and Yu G. Stroganov, {\it J. Stat. Phys.} {\bf 59}, (1990) 799.
\bibitem{vG} G. von Gehlen and V. Rittenberg, {\it Nucl. Phys.} {\bf B 257}, (1985) 351.
\bibitem{alc}F.\,C. Alcaraz and A. Lima Santos, {\it Nucl. Phys.} {\bf B 275}, (1986) 436.
\bibitem{p} Yu. A. Bashilov and S.\,V. Pokrovsky, {\it Commun. Math. Phys.} {\bf 76}, (1980) 129.
\bibitem{perk1} H. Au-Yang, B. M. McCoy, J.\,H.\,H. Perk, S. Tang and M.\,L. Yan, {\it Phys. Lett.} {\bf A 123}, (1987) 219.
\bibitem{perk2} B. M. McCoy, J. H. H. Perk, S. Tang and C. H. Sah, {\it Phys. Lett.} {\bf A 125}, (1987) 9.
\bibitem{37} R. J. Baxter, J. H. H. Perk and H. Au-Yang, {\it Phys. Lett.} {\bf A 128}, (1988) 138.
\bibitem{toda} S.\,N. Ruijsenaars, {\it Commun. Math. Phys.} {\bf 133}, (1990) 217.
\bibitem{45} R.\,J. Baxter, V.\,V. Bazhanov and J.\,H.\,H. Perk, {\it Int. J. Mod. Phys.} {\bf B 4}, (1990) 803.
\bibitem{43} G. Albertini, B.\,M. McCoy and J.\,H.\,H. Perk, {\it Adv. Stud. Pure Math.} {\bf 19}, (1989) 1.
\bibitem{49} R.\,J. Baxter, {\it J. Stat. Phys.} {\bf 120}, (2005) 1.
\bibitem{39} R.\,J. Baxter, {\it J. Stat. Phys.} {\bf 52}, (1988) 639.
\bibitem{Bax82}
R.\,J. Baxter, {\it Exactly Solved Models in Statistical Mechanics},
Academic Press, 1982.
\bibitem{cao1} J. Cao, W.\,-L. Yang, K. Shi and Y. Wang, {\it Phys. Rev. Lett.} {\bf 111}, (2013) 137201;
\\J. Cao, W.\,-L. Yang, K. Shi and Y. Wang, {\it Nucl. Phys.}  {\bf B 875},
(2013) 152;
\\J. Cao, S. Cui, W.\,-L. Yang, K. Shi and Y. Wang, {\it Nucl. Phys.} {\bf B 866}, (2014) 185;
\\J. Cao, W.\,-L. Yang, K. Shi and Y. Wang, {\it Nucl. Phys.}  {\bf B 877}, (2013) 152.
\bibitem{Wan15} Y. Wang, W.\,-L. Yang, J. Cao and K. Shi, {\it Off-Diagonal Bethe Ansatz for Exactly Solvable
Models}, Springer Press, 2015.
\bibitem{Kor93} V.\,E. Korepin, N.\,M. Bogoliubov and A.\,G. Izergin,
{\it Quantum Inverse Scattering Method and Correlation Function},
Cambridge University Press, 1993.
\bibitem{Cha94} V. Chari and A. Pressley, {\it A Guide to Quantum Groups}, Cambridge University Press, 1994.
\bibitem{Bax04} R.\,J. Baxter, {\it J. Stat. Phys.} {\bf 117}, (2004) 1.
\bibitem{Geh06} G. von Gehlen, N. Iorgov, S. Pakuliak and V. Shadura, {\it J. Phys.} {\bf A 39}, (2006) 7257.

\bibitem{Kul81} P. P. Kulish, N.\,Y. Reshetikhin and E.\,K. Sklyanin, {\it Lett. Math. Phys.} {\bf 5}, (1981) 393.
\bibitem{Ize81} A.\,G. Izergin and V.\,E. Korepin, {\it Sov. Phys. Doklady} {\bf 26}, (1981) 653; {\it Nucl. Phys.}
{\bf B 205}, (1982) 401.
\bibitem{Tar92} V.\,O. Tarasov, {\it Int. J. Mod. Phys.} {\bf A 7} (Suppl.1B), (1992) 963.
\bibitem{Kul82} P.\,P. Kulish and E.\,K. Sklyanin, {\it Lecture Notes in Physics} {\bf 151}, (1982) 61.
\bibitem{Kir87} A.\,N. Kirillov and N.\,Y. Reshetikhin, {\it J. Phys.} {\bf A 20}, (1987) 1565.
\bibitem{Bax82-1} R.\,J. Baxter and P.\,A. Pearce, {\it J. Phys.} {\bf A 15}, (1982) 897.
\bibitem{Baz89} V.\,V. Bazhanov and N.\,Yu. Reshetikhin, {\it Int. J. Mod. Phys.} {\bf A 4}, (1989) 115.
\bibitem{Bax89} R.\,J. Baxter, {\it Adv. Stud. Pure Math.} {\bf 19}, (1989) 95.

\bibitem{Nep04} R.\,I. Nepomechie,  {\it J. Phys.} {\bf A
34}, (2001) 9993; \\
R.\,I. Nepomechie, {\it Nucl. Phys.} {\bf B 622}, (2001) 615;  \\
R.\,I. Nepomechie, {\it J. Stat. Phys.} {\bf 111}, (2003) 1363;
\\
R.\,I. Nepomechie, {\it J. Phys.} {\bf A 37}, (2004) 433.
\bibitem{CLSW} J. Cao, H.\,-Q. Lin, K. Shi and Y. Wang,
{\it Nucl. Phys.} {\bf B 663}, (2003) 487.
\bibitem{Yan06} W.\,-L. Yang, R.\,I. Nepomechie and Y.\,-Z. Zhang, {\it Phys. Lett.} {\bf B
633}, (2006) 664.
\bibitem{Fra07} L. Frappat, R.\,I. Nepomechie and E.  Ragoucy, {\it J. Stat. Mech.}, (2007) P09008.

\bibitem{Cao14}J. Cao, W.\,-L. Yang, K. Shi and Y. Wang,
{\tt arXiv:1409.3646}.


\end{thebibliography}
\end{document}